# The Effectiveness of Business Process Visualisations: a Systematic Literature Review

Ed Overes[1] [0000-0002-4210-2342] and Flavia Maria Santoro[2][0000-0003-3421-1894]

**Abstract.** Business Process Visualisations (BPVs) have become indispensable tools for organisations seeking to enhance their operational efficiency, decision-making capabilities, and overall performance. The burgeoning interest in process modeling and tool development, coupled with the rise of data visualisation field, underscores the significant role of visual tools in leveraging human cognition. Unlike traditional models, data visualisation approaches graphics from a novel angle, emphasising the potency of visual representations. This review aims to integrate the domains of BPV and data visualisation to assess their combined influence on organisational effectiveness comprehensively. Through a meticulous analysis of existing literature, this study aims to amalgamate insights on BPVs impact from a data visualisation standpoint, advocating for a design philosophy that prioritises user engagement to bolster organisational outcomes. Additionally, our systematic review has unveiled promising avenues for future research, identifying underexplored variables that influence the efficacy of BPVs, thereby charting a path for forthcoming scholarly inquiries.

## 1. Introduction

Effective management and optimisation of business processes are critical for organisations to remain competitive in today's dynamic and complex business environment. To achieve operational excellence, organisations employ various techniques and tools, one of which is the use of Business Process Visualisations (BPVs) [23]. The human sense 'Sight' has a processing bandwidth that is thousands times stronger than our cognitive sense [71]. The picture superior effect is an important reason which made visualisation an integral part of many scientific areas, also in the field of business process analytics [4]. BPVs involve the representation of processes, workflows, and activities within an organisation, enabling stakeholders to gain a better understanding of the entire process [76,79], identify areas for improvement [1] and, understanding implications of user interaction [30]. Despite the growing adoption of BPVs, there is a need for a systematic evaluation of their effectiveness. Many visualisations in business process modelling still appeal on our cognitive sense rather to sight. Although a substantial number of papers has been published on various aspects of BPV, little research has been carried out to understand how to effectively apply graphical visualisation principles to business process analytics [4]. While some studies have demonstrated the positive impact of visualisations on organisational performance, others have reported mixed results or identified potential challenges and limitations. User interaction is open for further research [23] as is the role of interaction in visualisations [26], and visual feedback [23].

[1] Human-Data-Interaction Center, Zuyd University, Maastricht, The Netherlands; email: ed.overes@zuyd.nl
[2] Institute of Technology and Leadership, São Paulo, Brazil; email: flavia@inteli.edu.br

The goal of this paper is to provide a comprehensive understanding of the role and effectiveness of BPVs through a systematic literature review (SLR). The SLR aims to synthesise the existing evidence and provide insights for practitioners and researchers. In the next section, we first clarify some terms used throughout the paper and introduce the graphical perspective on effective visualisations. Next, we describe the methodological approach followed for this review. In Section 4, we report our findings, and in Section 5, we discuss other variables found which (potentially) have an impact on the effectiveness of visualisations. Finally, we end this review reporting our conclusions, discussion and final remarks.

## 2. Graphical Perspective on Effectiveness of Visualisations

In this review, we adopt the definition of visualisation as the transformation and presentation of data in a visual form to facilitate conception and understanding [40]. This definition was chosen because BPV aims to facilitate communication, strategic thinking and decision-making [1]. This means that visualisation has to communicate information clearly and efficiently but should not be limited to traditional forms like Bar, Line, Pie, Area, Stacked, Dual Axis, Combination, Donut, Scatter, Bubble, Tree Maps, Heat Maps, Geospatial Maps, Radar, Box Plots, Word Clouds, Waterfall, Parallel Coordinate, Funnel Charts and Grids [59]. Also, the seven genres of data storytelling [86] need to be mentioned since they are also linked with BPV [84]: 1) flow chart, 2) partitioned poster, 3) annotated charts/Infographic, 4) magazine style, 5) data comic, 6) video/animation, and 7) slide shows.

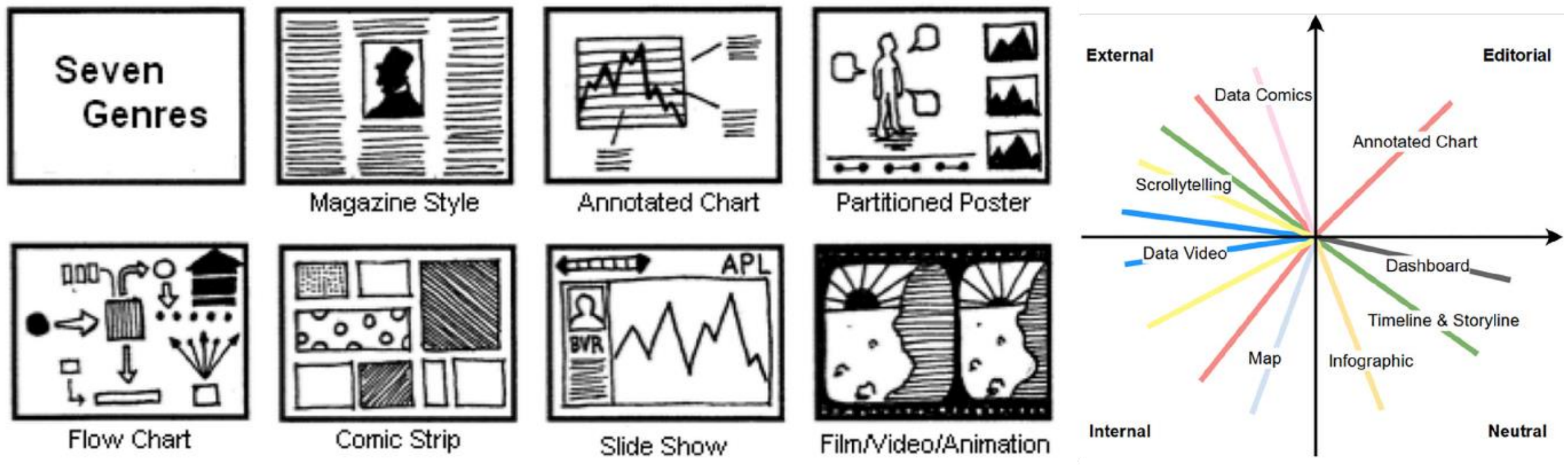

*Fig 1. Genres of narrative visualization proposed by Segel and Heer [86]*

In the visualisation literature, there are different views on the meaning of effective visualisation. Zhu identified *data-centric* ("effectiveness largely depends on the correspondence between the visualisation and the data it displays"), *task-centric* ("the effectiveness of visualisation is task specific") and *internal representation* views ("effectiveness of visualisation is about how accurate a data visualisation can be interpreted") [104]

An effective data-centric visualisation is one that displays the input data in a manner where the inherent structure of the data aligns seamlessly with the perceptual framework of the visualisation, ensuring clarity and insight into the data's underlying patterns and relationships [104] or when it maximises the Data/Ink Ratio [94]. This ratio principle suggests that effective visualisations maximise the amount of data displayed. However, there is no empirical evidence to confirm that a high Data/Ink Ratio directly contributes to more accurate interpretations or improved task efficiency [49,50]. Advocates of the data-centric view [93,94,97] argue that visualisations should be designed to be task-agnostic. Through user interactions, they can adapt to serve various tasks, thereby enhancing their utility and flexibility. In the field of visualisation, user interaction is defined as a dialogue between a user and the visualisation system over data [26]. Empirical psychological and computer-human interaction studies, however, tend to support the task-centric view when it comes to the effectiveness of visualisations [72] as task complexity has an impact on the effectiveness of visualisations [90], e.g. visualisations have little impact on task performance when task complexity is low.

The task-centric view focuses more on tasks than data when designing effective visualisations. In this view, effective visualisations are the ones that improve task efficiency and that designers should have a specific task in mind when they design the visualisation. Researchers who hold a task-centric perspective promote the adoption of a knowledge and task-based framework for the design and evaluation of visualisations. They argue that the effectiveness of a visualisation is inherently tied to the specific tasks it is intended to support [3,7,17,72].

The internal representation view focuses more on the user interaction [97] by stating that the effectiveness of visualisation depends on how accurate a data visualisation can be interpreted by the user [20,65]. In this review, we use the term 'user-centric' view instead of internal representation. From this perspective, effective visualisations are grounded in two fundamental principles for successful visualisation design: the Principle of Congruence, which emphasizes the alignment between the data structure and its graphical representation, and the Principle of Apprehension, which underscores the importance of creating visuals that can be easily and accurately interpreted by the viewer [96]. These principles indicate that the structure and content of a visualisation should match the desired mental representation (Congruence) and accurately perceived and comprehended by the user (Apprehension) [104].

Furthermore, a significant portion of the literature employs the term "data dashboard." We have incorporated these studies into our review, recognizing that dashboards serve as a cohesive collection of diverse, yet interconnected, data visualizations. Their purpose is to streamline the presentation of related information, making it more accessible and comprehensible to the viewer [92].

## 3. Methodological approach

This SLR followed a predefined protocol to ensure a transparent analysis of the literature. The selection process was carried out in several distinct phases (Figure 3), including the corresponding number of selected papers. The selection process is crucial as a comprehensive literature search is expected to include all relevant studies [44,47]. For this reason, the search should be conducted across four or more digital libraries [47]. In conjunction with an examination of existing BPV [23,103] and data visualisation [84] literature reviews, we have identified five relevant sources: (1) ACM Digital Library; (2) Emerald; (3) Ebsco; (4) IEEE Xplore; and, (5) Scopus. Based on our research questions, and consultations with BPM experts, we searched the title, abstract, and keywords, and formulated the following search string: **(“Business Process”) AND (Visualisation OR Visualization OR Dashboard) AND (Communication OR Understandability**

**OR Effectiveness OR Effects)**. First, the goal was not to conduct a review of process visualisations broadly but to specifically concentrate on BPVs. In preparation for this study, we observed that both 'Visualisation' and 'Dashboard' are terms commonly used to describe the enhancement of business process accessibility. To ensure no relevant articles were excluded based on terminology, we incorporated both terms in our search string. To explicitly associate 'Visualisation' and 'Dashboard' with business processes, we linked these terms using 'AND' in our search string. Additionally, to account for variations between British and American English spellings, we included 'visualisation' both with an 's' and with a 'z'. Second, our review aimed to specifically examine how visualisation impacts the effectiveness of business processes, intentionally excluding articles that focus solely on process modelling. This intention led us to augment our search string with 'AND Effectiveness.' During the pre-analysis phase, we realized that various terms were employed to describe the effectiveness of business processes. To capture a broad spectrum of relevant literature, we incorporated the most frequently used terms—'Communication,' 'Understandability,' 'Effectiveness,' and 'Effects' into our search criteria, ensuring an alignment of terminology. Although 'comprehension' was considered, we ultimately decided against its inclusion as it did not add unique value but rather duplicated existing search terms. Initially, we applied the search strings to our chosen databases without employing any filters to maintain a wide scope of potential articles.

We applied several exclusion criteria (EC) based on their format and publication details: EC1 – written language (Non-English articles); EC2 – type (not peer reviewed); EC3 – publication date (before 2009); EC4 – length (<=4 pages); EC5 – availability (non-full papers) (Figure 2).

| Criteria used | | |
|---|---|---|
| Exclusion criteria 1 | Written language | Exclude all Non-English articles |
| Exclusion criteria 2 | Type of article | The review will only include peer-reviewed scientific articles, and conference papers |
| Exclusion criteria 3 | Publication date | Exclude all papers from before 2009 |
| Exclusion criteria 4 | Length of article | Exclude short papers (<= 4 pages) |
| Exclusion criteria 5 | Availability | Exclude papers that are Non-full text |
| **Inclusion criteria 1** | The paper presents analysis of effects regarding visualization of process models | |
| **Inclusion criteria 2** | The paper presents a proposal to improve effectiveness about business process visualisations | |

*Fig. 2 Overview of the exclusion- and inclusion criteria*

For selection of the exclusion criteria we consulted relating reviews [23.103]. The year of 2009 was chosen as the starting point of this literature review because the first beta version of the BPMN 2.0 was released in that year [23]. While reading the articles we noticed several authors referring to the same papers written before 2009. In order to minimize the probability of leaving significant contributions behind, we noted down the specific articles referred to by three different authors, and based on the abstract, we included them in our review. EC2 and EC3 were defined to restrain the initial set of papers to those that were within the desired scope, while EC4 guaranteed a certain measure of quality. The criteria were applied manually.

The inclusion criteria were applied for selecting the papers that had their content related to this SLR topic, e.g. what defines effectiveness of business process visualisations and, as a consequence, how can this effectiveness be improved. The inclusion criteria are: IC1 - The paper presents analysis of effects regarding visualisation of process models; IC2 - The paper presents a proposal to improve effectiveness about BPVs. The abstracts of the selected articles were read and analysed by both authors and classified as 'Include' or 'Exclude'. If at least one of the two readers classified the paper as 'Include', the article was included in the review. At the same time each of the readers indicated whether the articles was included based on IC1, IC2 or both.

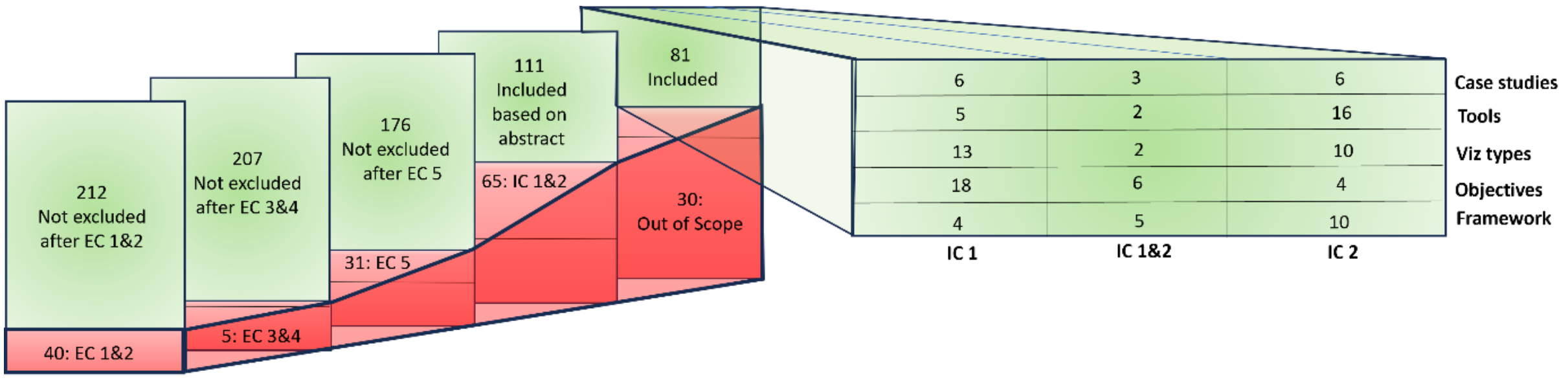

*Fig. 3 The results after applying the ex- (EC) and inclusion criteria (IC)*

The total amount of papers after applying each of the ECs and ICs is presented in Fig. 3. The whole selection process resulted in a set of 252 papers. Upon reviewing the abstracts, we identified multiple recurring themes. Consequently, we have categorized the articles into one or more groups, enabling authors to systematically explore the papers. In the appendix, we present an overview of how articles were classified. Finally, all included articles were read, and we came across 30 articles which appeared to fall out-of-scope of this review and were removed from the list after a cross-check with the other author. In Fig. 4 an overview of the segmentation is given of the articles read for this review.

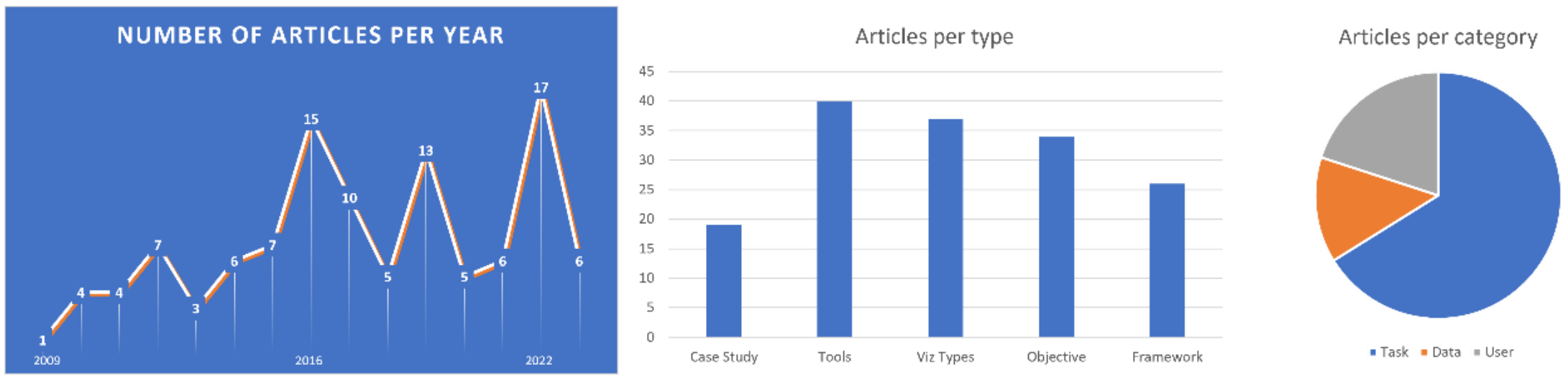

*Fig. 4. Review statistics with (a) number of articles per year, (b) number of articles per visualisation type, and (c) number of articles per category*

# 4. Business Process Visualisation Effectiveness

## 4.1 BPV frameworks used

In the selected articles, several frameworks were presented. The Visual Literacy Continuum distinguishes three different perspectives on BPV frameworks: *visual thinking, visual learning*, and *visual communicating* [85], which correspond with meanings for process design as a form of expression, a tool for thinking, and an aesthetic concept [8,9]. Visual thinking relates to the task-centric view. It is described as a process where information could be transformed into pictures to make standard modelling languages suitable for communicating procedures and recognising recurring structures [8]. The aim is not only to observe and analyse the visual, but also use it in decision making and communication. Other authors developed frameworks in line with 'visual thinking' and applied them in a visualising process mining outputs [88] or process performance comparisons [37,57,61,100]. In terms of effectiveness of BPV, it is necessary to examine how these frameworks aid analysts in identifying and analysing improvement opportunities [54]. Although the visual thinking perspective is widely used, these formal notations often do not describe the many exceptions users face when applying these procedures [8]. Referring to visualisations as tool of thinking, it is important to distinguish the iterative steps taken in a process visualisation [81].

Visual communication relates to the data-centric view. It uses symbols to share information [85] and refers to 'form of expression' [8] and 'syntax' [36], for a straight-forward model of conceptual data structures (data model) and the description framework for data visualisation (feature signature model). The goal is to develop a pattern language for visualising the spectrum of business processes to solve the artistic 'irritations' within organisations of formal modelling languages and notations [9]. The main benefit is improving organisational learning [6].

Visual learning is developing a concept through process abstraction, thus it is most connected with 'effectiveness' as it is related to understanding the message of the visualisations by users and finding a semantic fit to the data [36]. Effectively visualising semantics can be enhanced by mapping various data types (quantitative, ordinal, and nominal) to visual variables like position, size, hue, colour, shape, texture, and orientation [36].

The user level would not be complete without mentioning frameworks developed for dashboard visualisation. Finding the right balance between making visualisations attractive to get the readers' attention and reducing complexity [92,95] was considered important in the understandability of visualisations [15,29]. Dashboards are used to keep an overview of the BP performance, but most dashboards fail to communicate critical data efficiently and effectively due to glitzy, flashy displays [15]. Due to the unstable nature of BP, Bayesian networks are proposed to support process visualisation and include uncertainty-related knowledge [87]. A framework providing object annotations and representations to integrate both conceptual and physical modelling is proposed for an easy communication of a model between a practitioner and viewer [13].

## 4.2 Objectives of visualisations of processes

The effectiveness of process visualisations is gauged by the degree to which the visualisation achieves its intended objective. This underscores the importance of aligning the choice of visualisation with the specific goal in mind. Objectives for visualisations commonly fall into three categories: *explanation, exploitation*, and *exploration* [98]. Explanation focuses on displaying data for the sake of understanding (data-centric view), exploitation builds upon existing capabilities and opportunities (task-centric view), and exploration concentrates on developing new capabilities and opportunities (user-centric view). The exploitive objectives were dominant in the articles read, for example by looking at the effects of adding visuals (graphs) to data captured in tables [40], using routinised modelling metrics to emphasise exploitative activities [19]. Software solutions (Celonis, MPM, Minit) mostly focus on exploitation [54].

We found four perspectives for enhancing process models with performance data: 'In time-related process behaviour' [4,13,36], 'human resource behaviour over time' [61,73], 'multi-perspective process visualisation of process variants' [74], and 'predictive and prescriptive process monitoring' [21,24,55,101]. An explanatory approach uses a tree structure to elucidate hierarchical relationships, facilitating an overarching perspective when displaying multidimensional data [105]. It diversifies the methodologies for visualising process models, aiming for a holistic view of processes. It advocates for the use of less rigid modeling languages and notations, allowing for a more concrete and less abstract visual thinking process [9]. This recommendation is grounded in the contemporary practices observed in the field of business process modeling [52,53].

### 4.3 Visualisation types used in BP visualisations

In Fig. 5, we categorised the process visualisations used in the articles read, according to the distinctive views in paragraphs 2 and 4.2. Task-centric visualisations are dominant when visualising business processes. BPMN and Petrinets, and all plug-ins and add-ons fall in this category.

| | | EXPLANATORY | | EXPLOITORY | | | EXPLORATORY | |
|---|---|---|---|---|---|---|---|---|
| TASK | EC 1 | 15,58,60 | 79 | 79 | 4,21,24,28,54,75,83,102 | 89 | 89 | |
| | 1&2 | 9,34 | | | 55,81 | 31,88 | 31,88 | 1,14 |
| | EC 2 | 35,59,73 | 41 61 84 | 41 61 84 | 2,18,36,37,90,92,96 | 51 | 51 | 10,16,57,70 |
| DATA | EC 1 | 27,45,64 | | | 63,103 | 12 | 12 | |
| | EC 2 | 39,101,106 | | | 8,74 | | | |
| USER | EC 1 | 22,42,66 | 40 | 40 | 30,56,69 | 19,78 | 19,78 | 62,76 |
| | EC 2 | | 48 | 48 | 5,33,38 | | | 32 |

*Fig. 5 Categorisation of the different papers for this review*

In an empirical evaluation of BPMN, it was found that all patterns (Time, Cost, Path, Colour, History and Information) contributed to the understandability but were hard to apply on different managerial levels [60].

| Evaluation Pattern | Purpose | Constructs |
|---|---|---|
| Time Pattern | Analyse the performance of resources and activities with respect to time | Swimlanes, activities |
| Cost Pattern | Analyse the performance from the cost perspective like material and resources | Swimlanes, activities, colours |
| Path Pattern | To understand the activities which will be fruitful | Edges, activities, colours |
| Colours Pattern | To represent which activities are distinct in a process | Activities, swimlanes, connecting objects |
| History Pattern | To understand which activities are frequently executed in process | Edges, thickness, activities |
| Information Pattern | To provide further information along business process models | Gateway with rules, contents |

*Fig. 6 Visual patterns to facilitate the understanding of business processes; from [60]*

For presenting performance over time, three visualisations (flow- and time series charts, diagrams, and tables) simultaneously, were effective [70]. To enable interaction with the model to transition from 'as-is' to 'to-be' scenarios, the use of Petrinets for dynamic business process modelling [28] and transition flow diagrams consisting of geographical maps, and modified Sankey charts [101] were suggested. Task-centric visuals can be improved via text, graphical styling or tagging, with a potential conflict between aesthetics and modelling effectiveness [80]. Readability can be significantly improved by applying three distinct visual cues to visual tagging: hatching, shadowing, and the use of sketchy lines [101].

Heatmaps, timelines, treemaps and node-link diagrams are data-centric visualisations which have proven to be effective ways to overview a large number of data and find potential operational problems [5,45] when a multi hue map from the HSL colour system is applied [102]. To enhance the understanding of process executions, transforming event logs into on customisable, animated process maps is proposed as they provide a more intuitive understanding of activities and their performance over time [57]. To visualise and manage complex production workflows more effectively, an extension of BPMN with task scheduling and inventory management is described [91]. Investigations revealed that selecting a particular visualisation format did not significantly enhance decision-making accuracy but only partially contributed to increasing users' confidence in their decisions [62].

We concluded that effectiveness of business processes models and diagrams is improved by adding (1) icons for semantic meaning to symbols [33], (2) sequence of activities and dimensions [24] (3) graphical annotations [24,83] and (4) dimensions for data objects [39].

### 4.4 Tools and languages

In this paragraph several tools will be introduced, mainly if they give more clarification or introduce new perspectives on ways to visualise business processes. This paragraph is by no means a complete overview of the existing visualisation tools, it merely represents tools we encountered while reading the articles for this review and how they were used to visualise specific processes.

The four tasks that should be supported by modelling tools are Communication, Simulation, Creation/Modification and Verification [14], whereas user interface development, process definition interface and business process logic are the three parts of the business process visualisation tool requirements [18]. BPM tools should have six standard user design features for the development of advanced, interactive business process visualisations to facilitate more effective and efficient process design: Zoom, Filtering, Relationship Representation, Collaboration, Context and Direct Manipulation [14].

The majority of visualisation tools are developed with designers in mind, rather than end-users. However, the true measure of a visualisation's effectiveness lies in its ability to enhance communication, strategic thinking, and decision-making among users [1]. Many authors advocate to have a user-centric view [9,13,40,41,48,51,52,60,89,102] and process visualisations should be tailored to the specific needs of the stakeholder [51], with a focus on the human and ‘soft’ aspects like culture, organisational and political factors [9]. Many tools use variables (colour, shadow) for the sake of aestatics but not necessary meaningful for a given notation, i.e., colours are used to represent various purposes which makes it less suitable to distinguish [89].

Another dimension to consider is that users in operational roles typically possess extensive operational knowledge but may lack analytical or visualisation skills. This gap can make it challenging for them to identify potential issues within visualisations [102]. Although all forms of graphical styling could potentially contribute to the understandability, they were hard to be applied on different managerial levels. Time and Colour patterns were found useful as these patterns are providing decision support and are applicable in different areas, whereas other patterns were not [60]. Among the user-centric approaches, the DDCAL algorithm is particularly useful for visualising data on maps and business process models [63]. Annotated transition systems is a tool for identifying statistically significant differences between process variants [10]. To facilitate dynamic visualisation of the sequence of valid business rules at any given moment, ProM.1 was proposed [16]. Visual PPInot adds a technique to support the comprehension of PPI [80]. They enhance business process flow charts with PPI information by using a combination of small icons, name and ruler. The icons used, whether it indicates Time, Count, State, Condition, or Data are as suggested by the Physics of Notation principle of semiotic clarity [80].

OnLine Analysing Processing aims to improve the understanding and insight of decision-makers in an intuitive and multi-dimensional way [105].

### 4.5 Case studies

We only found a limited number of publications to be classified as cases studies and there was hardly any focus on the effectiveness of the visualisations used.

One study found that visualising data via compound timelines improved functionality (correctness and consistency), efficiency (completion time and user performance), and acceptance by the users [84], while others found positive effects on decision quality [39] but no impact on confidence in decision making [40]. Nearly all case studies examining the effectiveness of process visualisations concentrated on the users' understanding of the process and a need for a flexible, role-specific design [66].

Established BPM visualisations and analysis techniques show similar results in terms of comprehensibility, expressiveness, and user confidence compared to other techniques [58]. Case studies applying Symbol, Colour, Texture, Text, and Edge Pattern visualisations to an existing business process [35] found that Colour was best suited for gaining a quick overview [35], and regarded useful for providing decision support [102] but not to represent various purposes as it becomes less suitable to recognize and distinguish [89]. Internal stakeholders confirmed that Colour is best suited for obtaining a quick overview and Text for in-detail analysis of compliance states [35].

Text and Symbol should be preferred for an in-depth analysis of a business process as both visualisation types allow a easier mental association. Texture seems to be valued for the ability to convey information, but not for readability [35].

Ranking-based visualisations proofed to create a better understanding efficiency, and needed less perceived and partially less objective mental effort to understand inconsistencies than text-based tables [22]. Another study found that the more complex the tree visualisations became, the less effective it became in task completion [43].

## 5. User Perspective on BPV Effectiveness

Several additional variables emerged from the articles read, which might have an impact on the effectiveness of process visualisations. All these variables lead back to the user, as not all users have the same characteristics. In this paragraph we will elaborate on these variables.

Terms which were mentioned to measure effectiveness from a user perspective were comprehension and understandability. Comprehensibility is the most important influencial factor in the assessment of a model's overall quality, outranking completeness, correctness, simplicity, and flexibility [30]. Several authors defined understandability in the context of business process modelling [11,25,67,68,77,78]. For this review, we applied this definition to visualisation: the degree to which information contained in a business process visualisation can be easily understood by a user [77].

### 5.1 Visual Literacy and Visual Experience

Visual literacy refers to the competence of users to understand visualisations, based on education and experience in reading visuals. Various authors have noted differences in comprehension attributable to experience [9,30,66,78,82,102], so the reader's level of experience plays a crucial role in designing visualisations. A lack of effectiveness with users results in difficulty to find potential problems in used visualisations [66,102]. Syntax highlighting with colours is positively related to novices' model comprehension, but not to experts' understanding [78]. BPMN diagrams only lead to a better comprehension of diagrams at experienced users [82]. One author indicated that aesthetic visualisation would require some level of visual literacy [9].

When no statistically significant evidence of the impact of modelling experience on comprehension was found [42], it was suggested that this may be due to maintaining experience levels constant in research design [30]. Users prefer BPMN diagrams when trying to understand a business process [32], but this leads only to a better comprehension at experienced users [82].

Finally there seems no impact of special ability on visual preference, whereas both task difficulty and cognitive style impact the choice of visualisation format [62].

### 5.2 Culture, Media and Comprehension

Some authors reported cultural influences on the preference and comprehension of visualisations and its elements [27,34,43,56]. Colour coding was perceived differently in Asian compared to Western cultures [56]. Left-to-right flow direction was assumed to be preferred but no evidence was found [34]. Studies about the BPV in an African context assumed that illiteracy would be the explanation for preference for a visualisation form, which was was supported [12,42]. The cultural context on the effectiveness of BPVs was found in a study in an Eastern European academic context [27]. All these studies indicate a potential cultural influence on the comprehension of business process visualisations, but no study was found which studied these effects.

Additionally, one study revealed strong evidence that participants find models on paper easier to understand and more useful as it would take less effort in information-seeking in their specific setting [95]. This finding is in line with those of several other studies [64,68,69,75].

### 5.4 Gestalt and Cognitive Load Theory

For understanding the comprehension and effectiveness of visualisations, several authors refer to Gestalt theory [30,35,99] as it studies human perceptions. Gestalt theory deals with principles associated with how humans perceive whole figures instead of simpler, separate elements [99]. Gestalt theory emphasizes that structures have specific properties that are different from the sum of their individual parts. As a result, the attributes of the whole are not deducible from analysis of the parts in isolation [30]. The main concern of these articles is how to incorporate Gestalt theory in model design to make it easier for humans to recognize related elements as belonging together.

Cognitive Load Theory offers a foundation for enhancing learning and understanding in the realm of process model comprehension [90]. It posits that for effective comprehension, models must avoid being overly complex or burdensome [30]. Accordingly, visualisations play a crucial role in safeguarding readers against extraneous information by employing modularisation techniques to streamline the presentation of data [78,95]. Other work supporting the cognitive load theory mentions that well-designed dashboards involves prioritisation of the metrics [2]. The cognitive limitations of users should be taken into account when designing process visualisations [19]. The work of Clark is also a warning for potential manipulation by designers [19].

Comprehension is also affected by the extraneous cognitive loads, which means that visualising business process should take these extraneous cognitive loads into consideration [13,46]. In order to prevent unnecessary extraneous cognitive load Moody identified nine principles for designing notations: semiotic clarity, graphic economy, perceptual discriminability, visual expressiveness, dual coding, semantic transparency, cognitive fit, complexity management, and cognitive integration [31]. The practical significance of these principles are an improved end user understanding of visualisations, whereas the theoretical significance is that it provides important design implications as it gives insight into the effects of complexity on understanding of visuals.

## 6. Discussion and final remarks

This paper provided an extensive overview of the effectiveness of BPV from the perspective of data visualisation. Data visualisation and data storytelling are emerging fields for exploring and making data accessible. We used the lessons learned from these fields regarding visual effectiveness and connected it with the literature on BPVs. It opened new ways of looking at BPVs, explored ways to improve the effectiveness and discovered new promising directions for further research. We recommend process designers to not primarily have a task-centric view when designing new BPVs, but also have an eye for the user. This user-centric design takes into account the cognitive preferences and limitations of users, the cultural background and variety of users, as well as the (media) context in which visualisations are used by users.

Selecting relevant papers poses a challenge for all systematic literature reviews. For our study on the effectiveness of BPV, we sourced research papers from renowned, high-quality sources within the BPV field. Despite careful selection of sources and keywords, we acknowledge that our database may not encompass all pertinent studies, particularly excluding those focused on process modeling or mining. Our search criteria, informed by expert discussions and preliminary studies, were deliberately chosen to explore certain aspects of BPV, potentially overlooking others. This has led us to identify various avenues for future research, notably the need for more user-centric visualisations. We recognised variables related to user experience that affect the comprehension and effectiveness of visualisations, but found limited research on the influence of visual literacy, culture, cognitive style preferences, and cognitive load on users' understanding of visuals, indicating a need for further investigation in these areas.

**Appendix:** *Classification of articles (continued)(the articles in red have been found in several databases)*

| author | year | | IC1 | IC2 | Case Stu | Tools | Viz Types | Objective | Framework | 1 Explanation | Task/Data/User |
|---|---|---|---|---|---|---|---|---|---|---|---|
| Abubakre, Mumin and Fayoumi, A | 2021 | | X | X | X | | | | | 3 | T |
| Ahmed, Ali and Page, John and O | 2020 | | | X | X | | | | | 1 | T |
| Allio, Michael K | 2012 | | | X | | | X | X | | 2 | T |
| Ardito, Lorenzo and Cerchione, Ro | 2022 | | | X | | | | | X | | |
| Arnaboldi, Michela and Robbiani, A | 2021 | | | X | | | | | X | | |
| Bachhofner, Stefan and Kis, Isabe | 2017 | | X | | | | | X | | 2 | T |
| Barata, João and da Cunha, Paulo | 2015 | | | X | | X | | X | | | |
| Basole, R.C., Park, H., Gupta, M. | 2015 | | | X | X | | X | | | 2 | U |
| Basten, Haamann | 2018 | | | X | | | | | X | | |
| Belchior, Rafael and Guerreiro, Se | 2022 | | X | | X | X | | | | | |
| Blattmeier, Monika | 2022 | | X | X | | | | | X | 2 | D |
| Blattmeier, Monika | 2023 | | X | X | | | | X | X | 1 | T |
| Bolt, A., de Leoni, M. and van der | 2016 | | | X | | X | | | X | 3 | T |
| Brown, Ross A. | 2010 | | X | | | | | | X | 2&3 | T |
| Brown, Ross and Recker, Jan | 2009 | | X | X | | X | | X | | 3 | T |
| Bumblauskas, Daniel and Nold, He | 2017 | | X | | | | | | x | 1 | T |
| Burattin, Andrea and Cimitile, Mar | 2015 | | | X | | X | | | | 3 | T |
| Burgert, Oliver and Fink, Elena an | 2014 | | X | | | X | | | | | |
| Chen, Jiayao and Li, Xueqing and | 2010 | | | X | | X | X | | | 2 | T |
| Chongwatpol, Jongsawas | 2016 | | | X | X | | X | | | | |
| Clark, Bruce | 2021 | | X | | | | X | X | | 2&3 | U |
| Conforti, Raffaele and Fink, Sven a | 2016 | | X | | | | | X | | 2 | T |
| Corea, Carl and Nagel, Sabine an | 2020 | | X | | X | | X | | | 1 | U |
| da Costa, Janaina Mascarenhas H | 2019 | | | X | | X | | | | | |
| de Assis Santos, Leonardo and M | 2022 | | | X | | | | X | | | |
| de Leoni, M., Suriadi, S., ter Hofst | 2016 | | | X | | X | | | X | 3 | T |
| Delias, Pavlos | 2017 | | | X | | | X | | | | |
| del-Río-Ortega, Adela and Resina | 2019 | | X | X | | X | | | | 2 | T |
| Di Francescomarino, Chiara and G | 2022 | | X | | | | | X | | 2 | T |
| Dospan, Saida and Khrykova, Ana | 2022 | | X | | | | X | | | 1 | D |
| Durugbo, Christopher | 2014 | | X | X | X | | X | | X | | |
| Dymora, Pawel and Koryl, Maciej | 2019 | | X | | | X | X | | | | |
| Eaidgah, Youness and Maki, Alirez | 2016 | | | X | | | X | | X | | |
| Edu, Abeeku Sam | 2022 | | X | | X | X | X | | | | |
| Fadahunsi, Odunayo and Sathiyan | 2016 | | X | | X | | X | | | 2 | T |
| Falkenthal, Michael and Barzen, Jo | 2017 | | X | X | | X | X | | | | |
| Fernandes dos Santos, Nataly In{\ | 2023 | | X | | | X | | | | | |
| Figl, K. | 2017 | | X | | | | | X | | 2 | U |
| Figl, Kathrin and Mendling, Jan and | 2013 | | X | X | | | X | X | | 2&3 | T |
| Figl, Kathrin and Recker, Jan | 2016 | | | X | | X | | X | X | 2 | U |
| Figl, Kathrin and Recker, Jan | 2016 | | | X | | | | X | | 3 | U |
| Figl, Kathrin and Strembeck, Mark | 2015 | | X | X | X | | | X | | 1 | T |
| Free, Clinton and Qu, Sandy Q. | 2011 | | X | | | | X | | | | |
| Gall, Manuel and Rinderle-Ma, Ste | 2021 | | | X | X | | | | | 1 | T |
| Gulden, J. | 2016 | | | X | | | | | X | 2 | T |
| Guo, Hanwen and Brown, Ross ar | 2013 | | | X | | X | | | X | | |
| Guo, Liang and Sharma, Ruchi and | 2017 | | | X | | X | | | | | |
| Hamdam, Adli and Jusoh, Ruzita a | 2022 | | | X | | | | | X | 2 | T |
| Happa, Jassim and Agrafiotis, Ioa | 2021 | | | X | X | X | | | | 2 | U |
| Hipp, Markus and Mutschler, Bela | 2014 | | | X | | X | | | | 1 | D |
| Hirsch, Bernhard and Seubert, Anr | 2015 | | X | | X | | X | X | | 1&2 | U |
| Hull, Richard and Koschmider, Agr | 2016 | | | X | | X | | | | 1&2 | T |
| Iba~nez, Erick Leonel Garcia | 2022 | | X | | | | X | | | 1 | U |
| Kang, Bokyoung and Jung, Jae-Yo | 2011 | | | X | | | | X | | | |
| Katifori; C. Vassilakis; G. Lepoura | 2014 | | | X | X | | X | | | | |
| Keivanpour, Samira and Ait Kadi, I | 2018 | | X | | | X | X | | | 1 | D |
| Kerklaan, Leo | 2011 | | | X | | X | X | | | | |
| Khorasani, Mahyar and Loy, Jenni | 2022 | | X | | | | | | X | | |
| Kitsios, Fotis and Kamariotou, Ma | 2019 | | | X | | | | X | | | |
| Korczak; H. Dudycz; M. Dyczkows | 2012 | | | X | | | X | | | 1&2 | U |
| Krenn, Florian | 2018 | | | X | | | X | | X | 2&3 | T |
| Krogstie, John | 2012 | | | X | | X | | | | | |
| Krueakam, Suchapit and Suharitda | 2019 | | | X | | X | X | | | | |
| Kubrak, Kateryna and Milani, Fred | 2023 | | X | | | | X | X | | 2 | T |
| Kubrak, Kateryna and Milani, Fred | 2022 | | X | X | | | | X | X | 2 | T |
| Kummer, Tyge-F and Recker, Jan | 2016 | | X | | | | | X | | 2 | U |
| Lamest, Markus and Brady, Maire | 2019 | | X | | X | X | | | | 2 | T |
| Lavalle, Ana and Mate, Alejandro | 2023 | | X | | | X | | | | | |
| Leotta, Francesco and Mecella, M | 2020 | | X | | | X | | x | | 1 | T |
| Leyer, Michael and Aysolmaz, Bar | 2021 | | X | | | X | | | | | |
| Lindoo, Ed and Duncan, Denise | 2016 | | | X | | X | | X | | 1 | T |
| Lodhi, Azeem and Saake, Gunter | 2022 | | X | | | | X | | | 1 | T |
| Low, W.Z. and van der Aalst, W.M | 2017 | | | X | | | X | | | 1&2 | T |
| Luo, Wenhong | 2019 | | X | | | | X | X | | 3 | U |
| Lux, Marian and Rinderle-Ma, Stef | 2023 | | X | | | X | | | | 2 | D |
| Lv; Yanjun Cui; Jianyu Song; Zifen | 2017 | | X | | | X | | | | 1 | D |
| Maddah, Negin and Roghanian, Er | 2023 | | | X | X | X | | | | 2 | T |
| Madyatmadja, Evaristus and Dian | 2017 | | X | | X | | | | | 3 | U |
| Mamudu, Azumah and Bandara, W | 2023 | | X | | | X | | | | | |
| Martinez-Millana, Antonio and Lizo | 2019 | | X | | X | | X | | | 1 | U |
| McCartney, Steven and Fu, Na | 2022 | | X | | | X | | | | | |
| Mendling, Jan and Recker, Jan an | 2019 | | X | | | | | X | | | |
| Mendling, Jan and Strembeck, Ma | 2012 | | X | | | | | X | | 2 | U |
| Ng, Irene and Parry, Glenn and Sr | 2012 | | | X | | | | X | | | |
| Nguyen, Hoang and Dumas, Marlo | 2016 | | | X | | | X | | | 3 | T |
| Orji; E. Ukwandu; E. A. Obianuju; | 2022 | | X | | | | X | | | | |
| Pichler, Paul and Weber, Barbara | 2012 | | | X | | X | | | | 1 | T |
| Pika, Anastasiia and Wynn, Moe T | 2014 | | X | X | | | | X | | 2 | D |
| Pini, Azzurra and Brown, Ross and | 2015 | | X | | | | | X | | 2 | T |
| Recker, Jan and Reijers, Hajo A a | 2014 | | X | | | | | X | | 3 | U |
| Reijers, Hajo A and Mendling, Jan | 2010 | | X | | | | | X | | 2&3 | U |
| Reijers, Hajo A and Mendling, Jan | 2011 | | X | | | | | X | | 1&2 | T |
| Reijers, Hajo and Recker, Jan and | 2010 | | | X | | | | | X | | |
| Roam, Dan | 2013 | | | X | | | | | X | | |
| Rodrigues, Raphael De A and Bar | 2015 | | X | | | | | X | | 2 | T |
| Rosa, Leonardo Silva and Silva, T | 2022 | | | X | | X | | | | 1&2 | T |
| Shilnikova; S. A. Odinokov | 2020 | | X | X | | | X | | X | 2&3 | T |
| Sirgmets, Marit and Milani, Fredrik | 2018 | | X | | | | | | X | 2&3 | T |
| Störrle | 2019 | | | X | | X | X | | | 2 | T |
| Szelagowski, Marek and Biernacki | 2022 | | | X | | X | X | | | 2 | T |
| Tarhan, Ayca and Turetken, Oktay | 2016 | | X | X | | | | | X | | |
| Turetken, Oktay and Dikici, Ahmet | 2020 | | X | | | | X | | | | |
| Turetken, Oktay and Rompen, Tes | 2016 | | | X | | | | | X | 2 | T |
| Wagemans, Johan and Feldman, J | 2012 | | X | | | | | X | X | | |
| Wynn, Moe Thandar and Poppe, E | 2017 | | | X | | | | | X | 1 | D |
| Xu, Zhi and Zhang, Zhi and Gui, Y | 2019 | | X | | | | X | X | | 2 | T |
| Yagi, Sayaka and Tsuchikawa, Kir | 2019 | | X | | | | X | | | 2 | D |
| Zhi, Qiang and Zhou, Zhengshu | 2022 | | X | X | X | | | | | | |
| Zhou, Jianlong and Arshad, Syed Z | 2018 | | X | | | X | | | | | |
| Zou, Benyuan and You, Jinguo and | 2019 | | | X | | X | | | | 1 | D |